\documentclass[conference]{IEEEtran}

\IEEEoverridecommandlockouts   

\usepackage{cite}
\usepackage{url}
\usepackage{graphicx}

\usepackage{indentfirst}
\usepackage{array}
\usepackage{fancyhdr}
\usepackage{float}
\usepackage{booktabs}
\usepackage{multirow}
\usepackage{here}
\usepackage{comment}
\usepackage{siunitx}
\usepackage{hyperref}
\usepackage{algorithmicx, algorithm,algpseudocode}
\usepackage{amsmath,amssymb,amsfonts,bm,gensymb}
\usepackage{fontawesome5}
\usepackage{caption}
\usepackage{setspace}
\usepackage{xcolor}
\usepackage{svg}
\usepackage{color}

\usepackage{comment}
\usepackage[normalem]{ulem}
\useunder{\uline}{\ul}{}

\usepackage{subcaption}

\def\BibTeX{{\rm B\kern-.05em{\sc i\kern-.025em b}\kern-.08em
    T\kern-.1667em\lower.7ex\hbox{E}\kern-.125emX}}

\newcommand{\etal}{et~al.\ }
\newcommand{\eg}{e.\,g.,\ }
\newcommand{\ie}{i.\,e.,\ }

\usepackage{hyperref}

\begin{document}

\title{\huge Enhancing the Driver's Comprehension of ADS's System Limitations: An HMI Providing Request-to-Intervene\\Trigger and Reason Explanation}


\author{
Ryuji~Matsuo$^{1}$, Hailong~Liu$^{1}$, Toshihiro~Hiraoka$^{2}$, Takahiro~Wada$^{1}$
\thanks{
$^{1}$Ryuji~Matsuo, Hailong~Liu, and Takahiro~Wada are affiliated with the Graduate School of Science and Technology, Nara Institute of Science and Technology, 8916-5 Takayama-cho, Ikoma, Nara 630-0192, Japan. \faIcon[regular]{envelope}~:~{\tt\small \{matsuo.ryuji.mr4; liu.hailong; t.wada\}@is.naist.jp}
}
\thanks{
$^{2}$Toshihiro~Hiraoka is with the Mobility Research Division, Japan Automobile Research Institute (JARI), 1-1-30 Shibadaimon, Minato-ku, Tokyo, 105-0012, Japan. \faIcon[regular]{envelope}~:~{\tt\small thiraoka@jari.or.jp}
}
}

\maketitle

\IEEEpubidadjcol

\begin{abstract}
Level 3 automated driving systems (ADS) have attracted significant attention and are being commercialized.
A Level 3 ADS prompts the driver to take control by requesting to intervene (RtI) when its operational design domain (ODD) or system limitations are exceeded.
However, complex traffic situations can cause drivers to perceive multiple potential triggers of RtI simultaneously, causing hesitation or confusion during take-over.
Therefore, drivers need clearly understand the ADS's system limitations to ensure safe take-over. 
In this study, we propose a voice-based instructional HMI for providing RtI trigger cues and reason to help drivers understand ADS's system limitations.
The results of a between-group experiment using a driving simulator showed that incorporating effective trigger cues and reason into the RtI enabled drivers to comprehend the ADS's system limitations better.
Moreover, the vast majority of participants, instructed via the proposed method, could actively takeover control of the ADS in cases where RtI fails, thereby significantly reducing the probability of collisions.
Therefore, using our proposed method to continually enhance the driver's understanding of the system limitations of ADS is crucial for promoting safer and more effective user interactions with ADS in real-time.
\end{abstract}

\section{INTRODUCTION}

Automated driving systems (ADS) have attracted considerable attention in recent years. 
Level 3 ADS, referred to as conditional driving automation, is also steadily progressing towards commercialization.
Level 3 ADS triggers a request-to-intervene (RtI), requiring the driver to takeover control of the vehicle when it approaches the limit of operational design domain (ODD) or system limitations, as defined in~\cite{sae2018taxonomy}.
After the RtI is issued, the ADS will also disengage when the ODD or system limitations is exceeded.
Therefore, the driver's prompt response to the RtI is crucial for ensuring safety.

\subsection{Issues in take-over}

Generally, after the ADS issued the RtI, the driver took approximately 10 seconds to respond, followed by a stabilization period of approximately 35-40 seconds to regain vehicle control~\cite{merat2014transition, strand2014semi}.
This delay can be attributed to the driver's need to recognize the reasons for the intervention, regain situational awareness, and assume vehicle control. 
Additionally, the time taken to regain control after taking over is essential for the driver to fully adapt to vehicle dynamics, road conditions, and overall driving demands.

\begin{figure}[t]
  \centering
  \includegraphics[width=\linewidth]{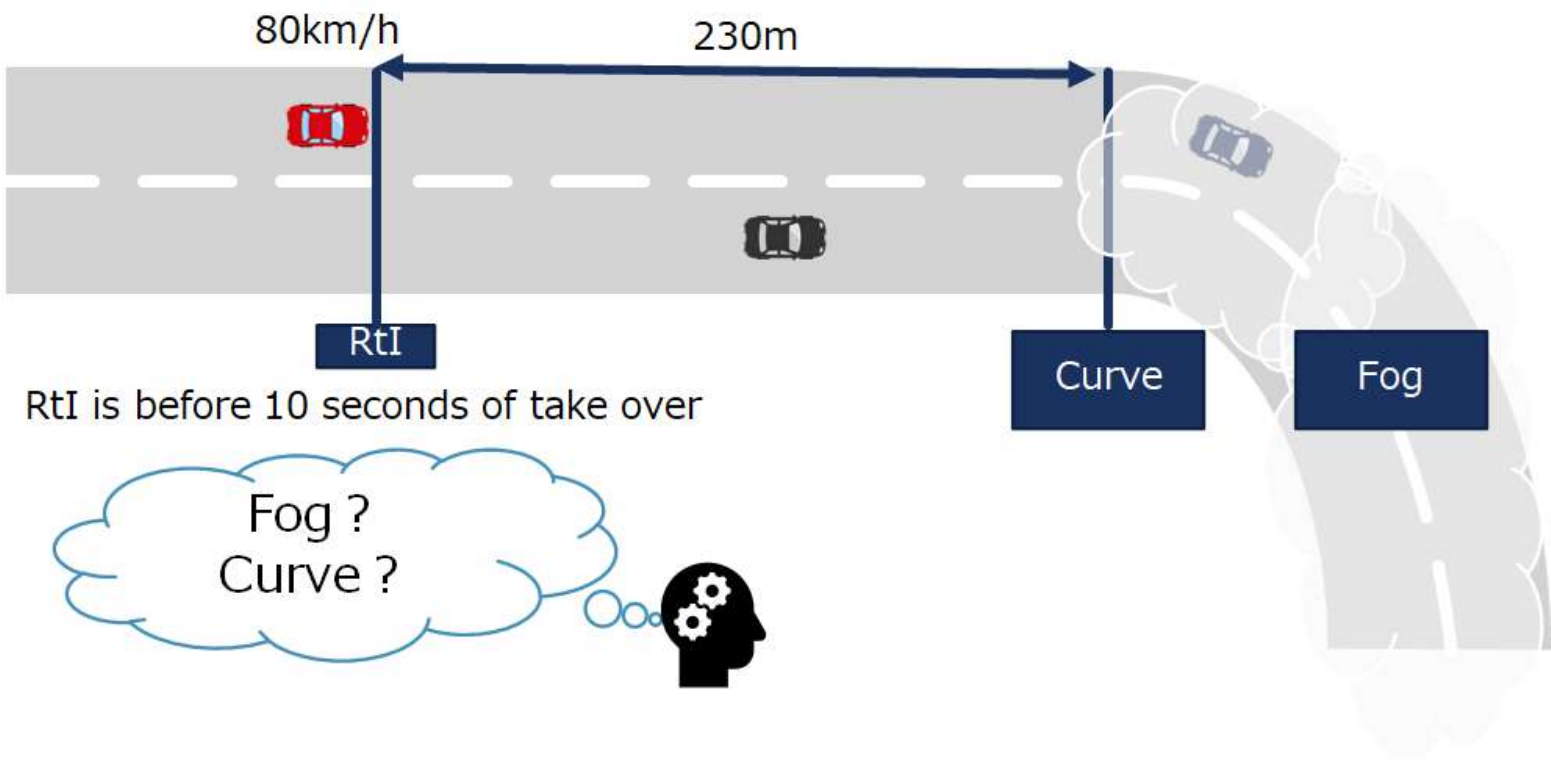}
  \vspace{-8mm}
  \caption{Illustration of a complex traffic scenario where the driver may encounter challenges in determining the cause of an RtI trigger.}
  \label{fig:scenario_ex}
  \vspace{-6mm}
\end{figure}

Moreover, the driver may perceive multiple potential triggers and hesitate or get lost entirely, even though the ADS issues an RtI for a single trigger in complex traffic situations. 
For example, Fig.~\ref{fig:scenario_ex} shows a complex and challenging traffic situation where foggy conditions on a curved road can cause drivers to perceive multiple triggers simultaneously. 
If the ADS issues an RtI in such a scenario, it will be challenging for the driver to determine whether the RtI is triggered specifically for the curve or because of the fog's presence. 
Moreover, overtrusting in the ADS may ultimately lead to accidents if the driver misunderstands the limitations of the ADS and misinterprets the triggers for RtI~\cite{liu2019overtrust}.



\subsection{Related works}

To address the above issues, the driver needs to 1) understand potential challenges the ADS may encounter, 2) recognize trigger the RtI protocol based on the surrounding conditions, and 3) take appropriate actions during the takeover.
This requires the driver to cognitively comprehend the limitations and capabilities of the ADS.
In essence, to safely take control of the vehicle during an RtI, the driver needs establish a precise mental model of the ADS.
The mental model represents the drivers' comprehension of the ADS mechanism and its limitations~\cite{liu2019driving,zhou2021does}.
For this, drivers naturally develop their mental model of ADS through repeated use, recognition, and interpretation of the system~\cite{staggers1993mental}. 
On the other hand, the pre-education can be use to develop a mental model of ADS quickly~\cite{zhou2021does, liu2021importance}.

Zhou~\etal focused on providing prior knowledge regarding the system limitations of the level 3 ADS and how it can influence driver intervention~\cite{zhou2021does}.
Their findings revealed that prior knowledge improved the success rate of take-over and reduced reaction time during the first experience with the RtI.
However, the learning effect of the original knowledge is diminished because drivers easily forget it over time after a few RtI experiences. 
Moreover, it can be challenging for drivers to identify the correct trigger and respond appropriately based on their prior knowledge when there are multiple possible triggers of the RtI from the driver's point of view. 

In addition, Wright~\etal proposed a method to inform drivers specific cues about upcoming hazards in order to shorten the time required for them to take over driving, and their experimental results showed that the risk of collision was reduced~\cite{wright2018effective}.
However, Wright~\etal did not discuss drivers' comprehension of the system limitations of ADS and did not show the effectiveness of the educational effect.

\subsection{Purpose and research questions}
\label{sec:RQ}

The purpose of this study is to propose a human-machine interface (HMI) designed to assist drivers in ensuring that they consistently establish an accurate mental model of the ADS, understand the system limitations precisely, and make correct inferences about triggers and reasons for RtI, particularly in scenarios where multiple triggers may appear ambiguous from the driver's perspective.

For this purpose, we propose a voice-based instruction HMI that consists of two components: 1) RtI trigger cues to assist drivers in rapidly enhancing their situational awareness, specifically to help drivers correctly identify the correct trigger for the RtI of an ADS among multiple potential triggers; 2) post-event reason explanations to help drivers learn about the system limitations of the ADS.
For example, in the scenario shown in Fig.~\ref{fig:scenario_ex}, the ADS provides an RtI trigger cue via voice, \eg, ``Thick fog, take over! Thick fog, take-over!'' 
In addition, after the driver turn ON the ADS after the take-over event, the ADS provides an RtI reason via voice, \eg, ``The reason for RtI is that it is difficult to recognize lanes and vehicles ahead in thick fog when it is impossible to see more than 40~m ahead.''

In this study, we conducted a between-group design experiment using a driving simulator to validate the proposed method's effectiveness.
We also compared its performance with a conventional RtI without trigger cues and reason explanations.
We addressed the following questions in the experiment:
\begin{description}
    \item[RQ~1:] Does the proposed RtI HMI, with a trigger cue and reason explanation, help the driver to correctly comprehend the system limitations of the ADS?
    \item[RQ~2:] If the driver correctly understands the system limitations of the ADS through repeated use of the proposed HMI, could it lead to the driver actively and promptly taking over the control during a take-over event even when the RtI fails?
    \item[RQ~3:] If the driver correctly understands the system limitations of the ADS through repeated use of the proposed HMI, could it lead to reduced accidents during a take-over event even when the RtI fails?
 \end{description}

\section{METHOD}

\subsection{Participants} \label{sec:participants}

For this experiment, we recruited 20 participants aged between 22 and 26 years.
They were randomly assigned to two groups which were: with trigger cue \& reason group (N=10) and without trigger cue group (N=10). 
All the participants had the driving licenses in Japan. 
Before the experiment, each participant provided informed consent and received a reward of 2,000 Japanese Yen in appreciation for the two hours of participation.

This experiment received approval from the Research Ethics Committee of Nara Institute of Science and Technology (No.~2022-I-56).

\subsection{Driving simulator which implemented a level 3 ADS}
\label{sec:DS}

\subsubsection{Equipment}
The experiment used a driving simulator based on the UC/win-Road driving simulator by Forum8 Inc., as shown in Fig.~\ref{fig:DS}.
The vehicle operation hardware included a \textit{Logitech G29} pedal set, and a \textit{SensoWheel SD-LC} force feedback steering wheel. 
The driving simulator displayed graphics on three 55-inch LED displays with resolutions of $1920 \times1080$ pixels each (see Fig.~\ref{fig:DS}).

\subsubsection{Level 3 automated driving system (ADS)}
\begin{figure}[t]
  \centering
  \includegraphics[width=0.9\linewidth]{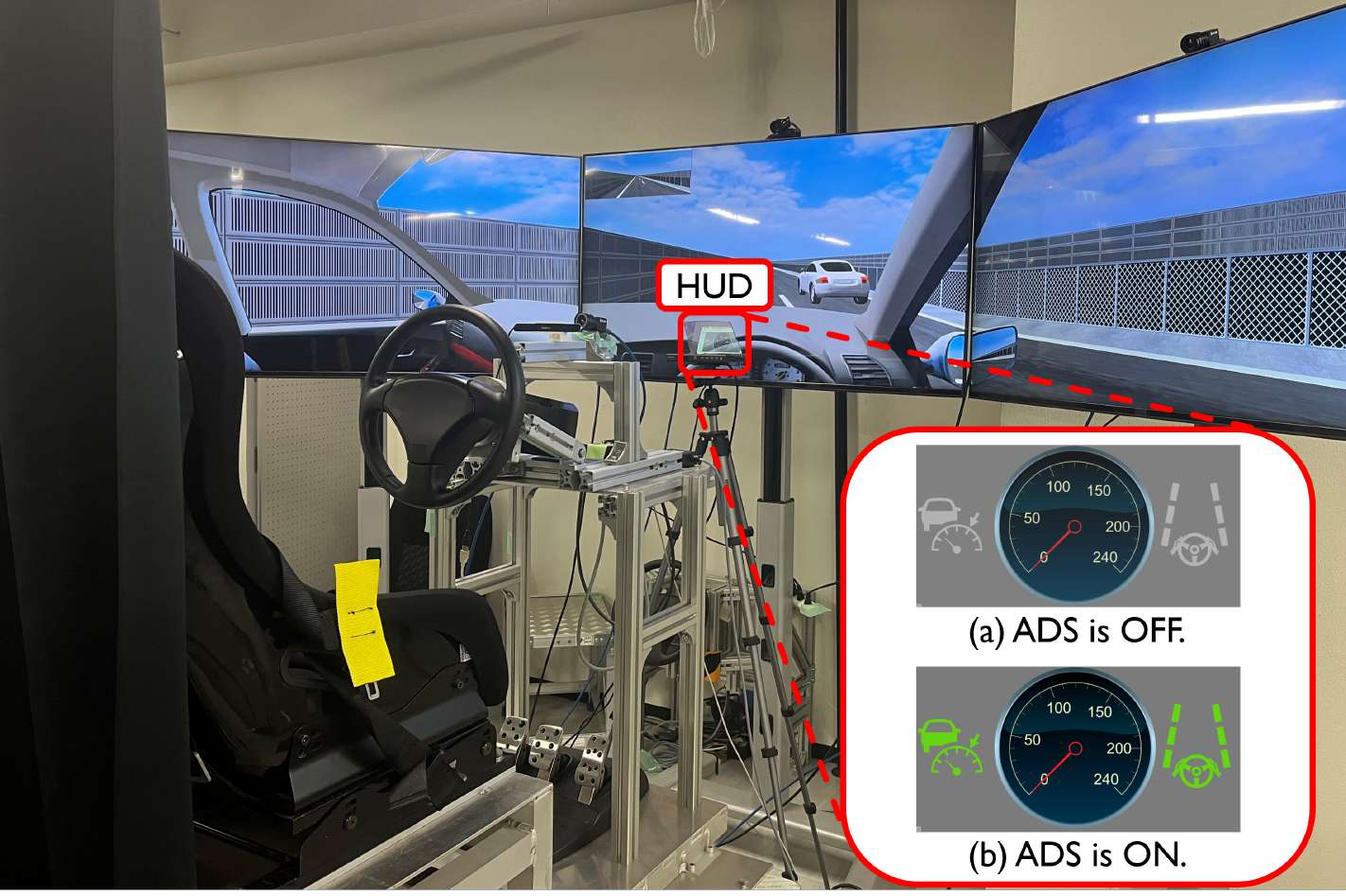}
  \caption{Driving simulator used in this experiment. A HUD shows the ADS status.}
  \label{fig:DS}
  \centering
  \includegraphics[width=\linewidth]{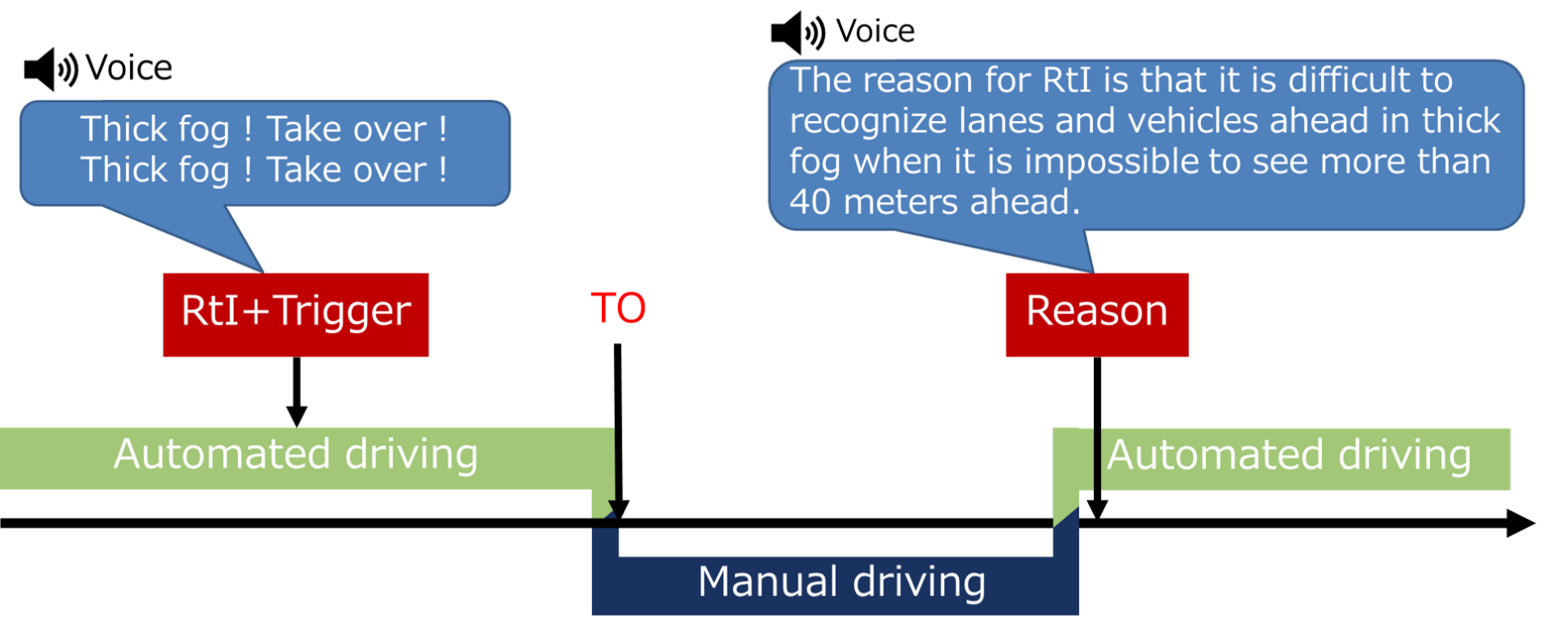}
  \caption{Proposed HMI that provides the driver information about the triggers and reasons for the RtI.}
  \vspace{-6mm}
  \label{fig:RtI_pro}

\end{figure}

The driving simulator reproduces an AV which has a level~3 ADS including an adaptive cruise control (ACC) system and a lane-keeping assistance (LKA) system. 
The ACC system automatically regulates the accelerator and brake to follow the preceding vehicle.
Besides, it maintains a speed of 80~km/h if there is no preceding vehicle.
The driver can override the ACC function during automated driving by pressing the accelerator pedals.
The LKA system automatically controls the steering wheel to keep the vehicle centered within the lane by detecting the lane markings. 
The driver can activate the ACC and LKA systems by pressing the clutch pedal during manual driving. 
In addition, the driver can deactivate the ACC and LKA systems by pressing the brake pedal or turning the steering wheel with a torque greater than 5~N torque.
As shown in Fig.~\ref{fig:DS}, the state of ADS, i.e., ON or OFF, is shown on a the head-up display (HUD).
In particular, the green and gray icons indicate that the ADS is ON and OFF, respectively.

\subsubsection{System limitations}
\label{sec:system_limitations}
We designed an experiment to incorporate two specific system limitations, i.e., the conditions for RtI to be issued:
1) Curve with a radius of less than 230~m,
2) Visibility range less than 40~m.
The RtI is triggered when any of the conditions exceed their thresholds.
Not that the ADS can function normally even in thick fog if the visual range remains above 40 m. 
In such cases, the RtI will not be triggered.

\begin{figure}[htbp]
  \centering
  \begin{subfigure}[b]{1\linewidth}
  \includegraphics[width=1\linewidth]{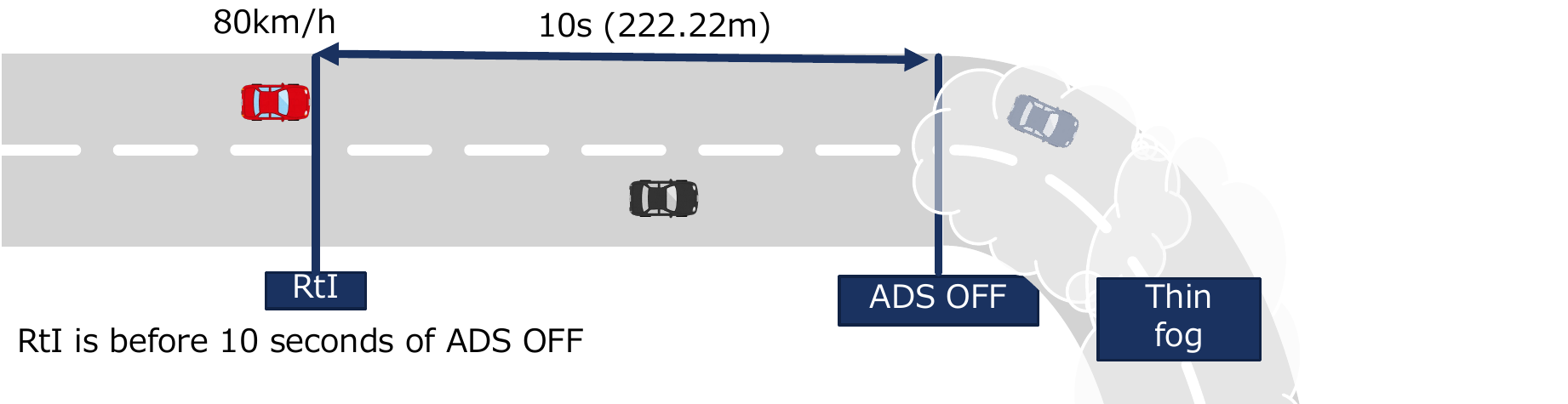}    \vspace{-5mm}
    \caption{A sharp curve road with thin fog.}
  \end{subfigure}\\ \vspace{1mm}
  \begin{subfigure}[b]{1\linewidth}
  \includegraphics[width=1\linewidth]{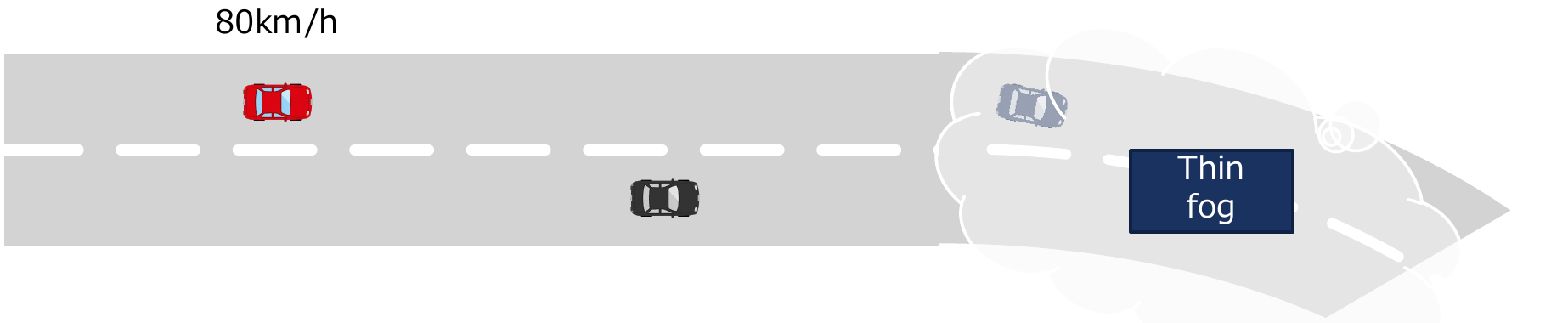}    \vspace{-5mm}
  \caption{A slight curve road with thin fog.}
  \end{subfigure}\\\vspace{1mm}
  \begin{subfigure}[b]{1\linewidth}
  \includegraphics[width=1\linewidth]{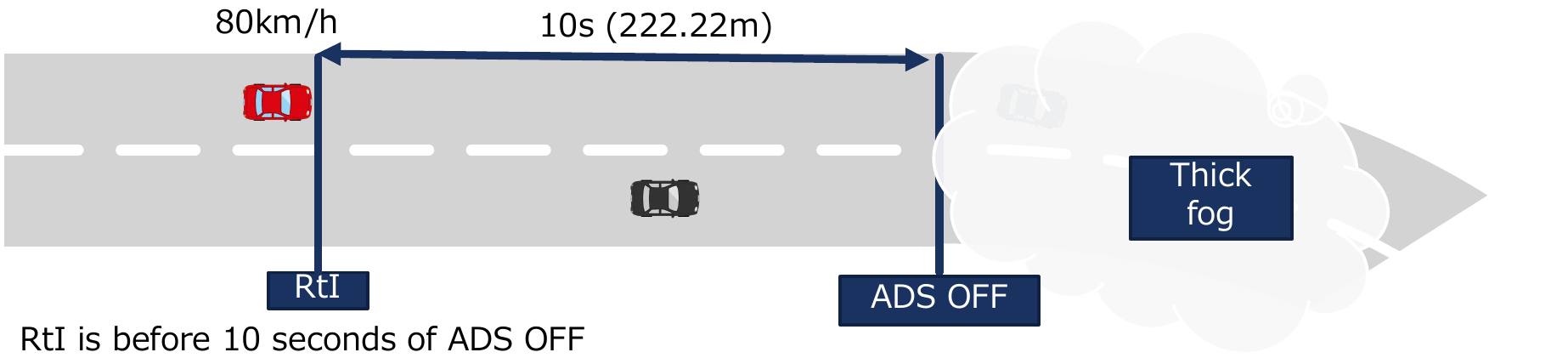}    \vspace{-5mm}
  \caption{A slight curve road with thick fog.}
  \end{subfigure}\\\vspace{1mm}
  \begin{subfigure}[b]{1\linewidth}
  \includegraphics[width=1\linewidth]{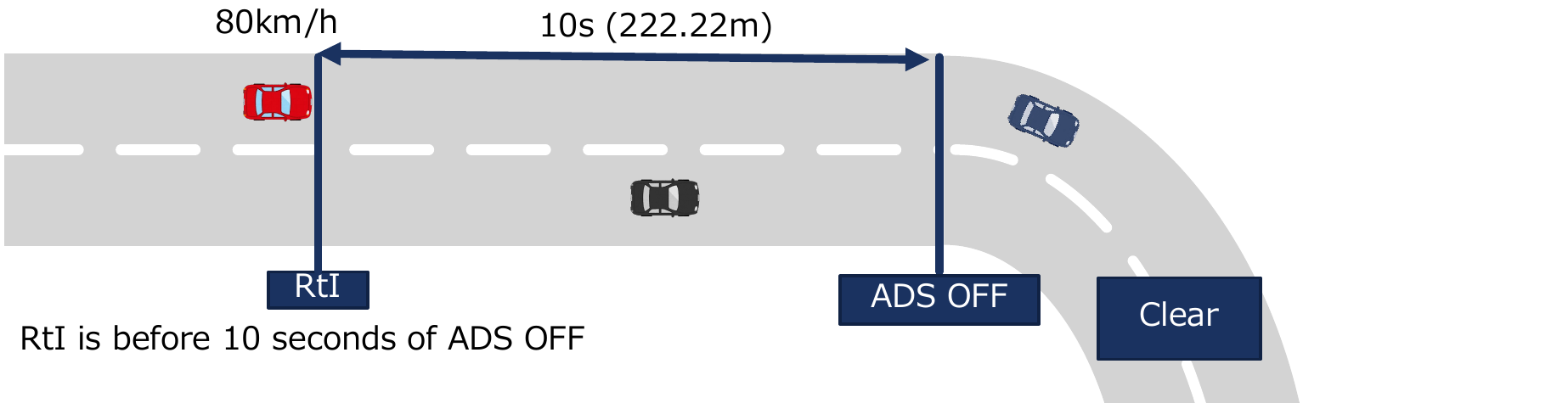}    \vspace{-5mm}
  \caption{A sharp curve road without fog.}
  \end{subfigure}\\\vspace{1mm}
  \begin{subfigure}[b]{1\linewidth}
  \includegraphics[width=1\linewidth]{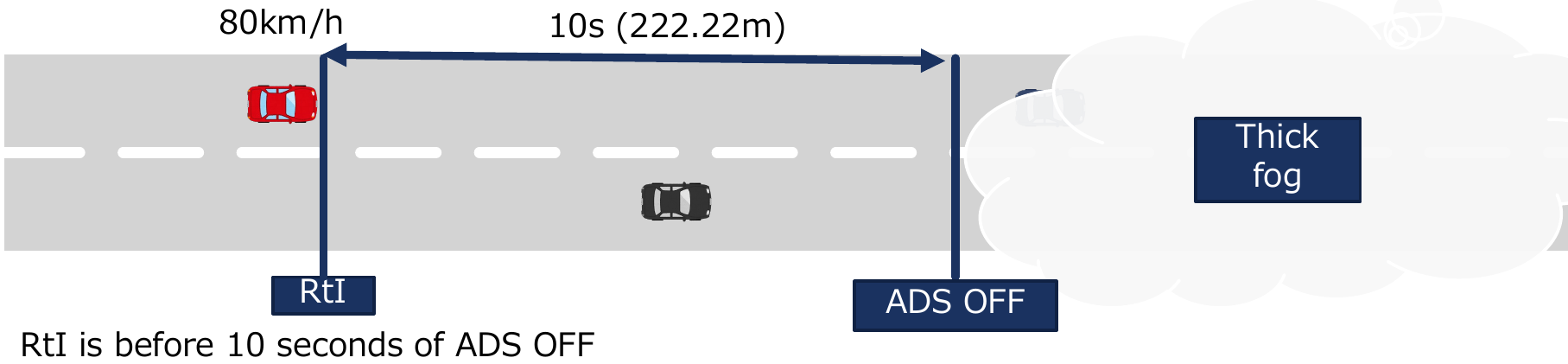}    \vspace{-5mm}
  \caption{A straight road with thick fog.}
  \end{subfigure}\\\vspace{1mm}
  \begin{subfigure}[b]{1\linewidth}
  \includegraphics[width=1\linewidth]{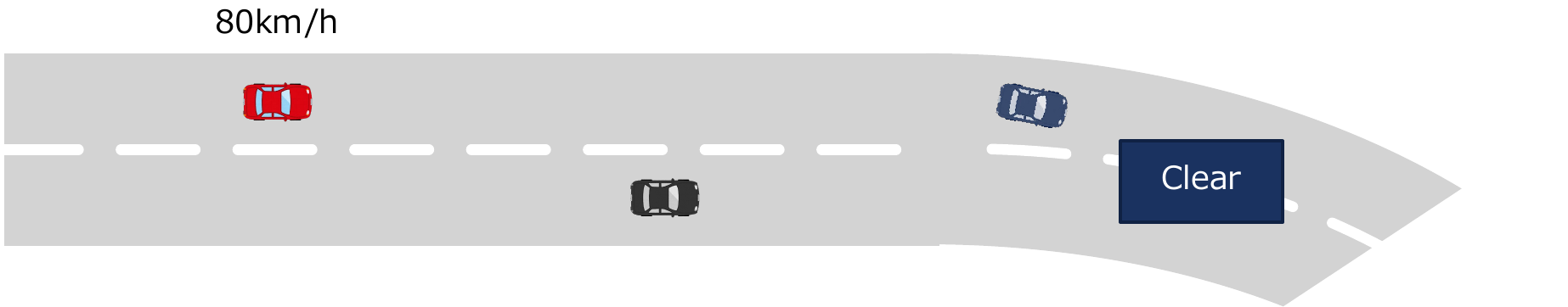}    \vspace{-5mm}
  \caption{A slight curve road without fog.}
  \end{subfigure}  \\\vspace{1mm}
  \begin{subfigure}[b]{1\linewidth}
  \includegraphics[width=1\linewidth]{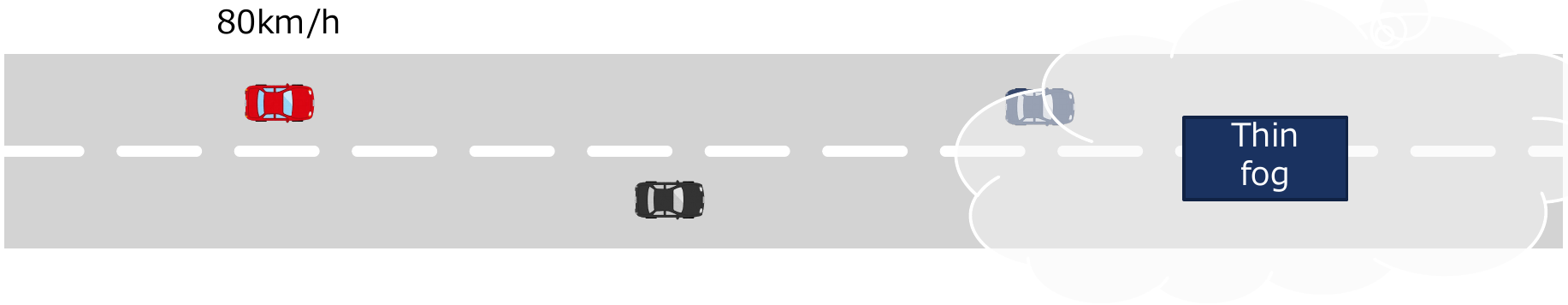}    \vspace{-5mm}
  \caption{A straight road with thin fog.}
  \end{subfigure}   \\\vspace{1mm}
   \begin{subfigure}[b]{1\linewidth}
  \includegraphics[width=1\linewidth]{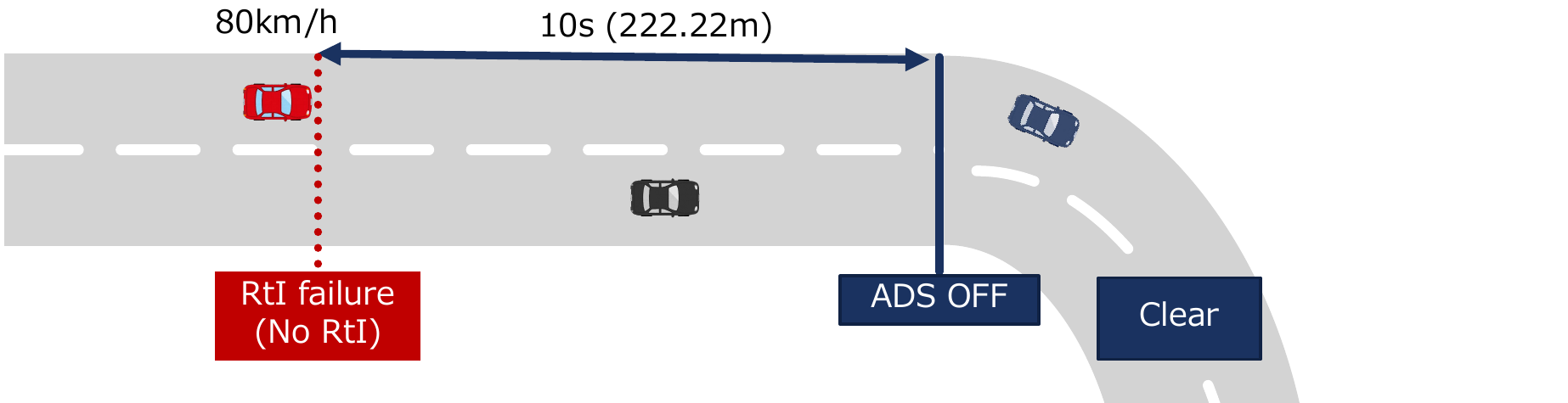}    \vspace{-5mm}
  \caption{A sharp curve road without fog but the RtI is failed.}
  \end{subfigure}  
  \caption{The various traffic scenarios used in the experiment, where the red car is the ego AV with a speed of 80 km/h. The scenarios (a), (b), and (c) are used in the phase I; (d) and (e) are  used in the phase II; and (f) and (g) are used in the phase III; (h) is used in the phase IV.}
  \label{fig:scenario}
    \vspace{-5mm}
\end{figure}

\subsection{Designs of RtI HMI}

As Fig.~\ref{fig:RtI_pro} shows, the RtI is designed to be triggered 10 seconds before the ADS exceeding the system limitations, referring to~\cite{merat2014transition}
As the ADS will be deactivated once it exceeds the system limitations, the driver needs to promptly and safely take over vehicle control.
To assist the driver during take-over, we propose a voice-based RtI HMI.
This is primarily because Petermeijer~\etal\cite{petermeijer2017driver} demonstrated that auditory and tactile take-over requests yielded faster reactions than visual take-over requests.

The proposed RtI HMI provides detailed information regarding the trigger cues and a post reason explanations.
In particular, the ADS issues the RtI with their corresponding voice cues from in-car speakers under the following conditions according to the system limitations: 
1) If a curve with a radius less than 230~m is detected, the ADS will issue the RtI with the voice cue: ``Sharp curve, take over! Sharp curve, take-over!''.
After the driver reactivates the ADS after the take-over event, it will issue the voice cue for the reason explanation as: ``The reason for RtI is that it is difficult to recognize lanes and vehicles ahead in a sharp curve with a radius less than 230~m.''
2) If the visibility range less than 40~m due to thick fog, the ADS will issue the RtI with the voice cue: ``Thick fog, take over! Thick fog, take-over!''.
After the driver reactivates the ADS after the take-over event, it will issue the voice cue for the reason explanation: ``The reason for RtI is that it is difficult to recognize lanes and vehicles ahead in thick fog when it is impossible to see more than 40~m ahead.''  

As a comparative method, we used a conventional voice-based RtI that issues a voice cue of ``Take over! Take over!'' without the trigger cue and the reason explanation.

\subsection{Scenarios}

\begin{table*}[ht]
\footnotesize
\setlength\tabcolsep{6pt}
  \caption{Experiment scenarios (conditions for RtI to be issued: a curve with R $<$ 230~m or the visibility range $<$ 40~m)}  
  \label{table:Scenario}
  \centering
  \begin{tabular}[H]{cccrcrr}
    \toprule
    Phases&\begin{tabular}[c]{@{}c@{}}Trial\\ No.\end{tabular}  & \begin{tabular}[c]{@{}c@{}}Scenario\\ shown in Fig.~\ref{fig:scenario}\end{tabular}    & \multicolumn{1}{c}{Visibility conditions} & Road conditions  &  \multicolumn{1}{c}{RtI}  &  \multicolumn{1}{c}{Trigger of RtI} \\
   \midrule
    I & 1 & (a)  &  Thin fog (Visibility range $\geq$ 40 m)  & Sharp curve (R=180~m)  & Issued  & Sharp curve \\
    I & 2 & (b)  &  Thin fog (Visibility range $\geq$ 40 m)  & Slight curve (R=280~m)  & Not issued  & None \\
    I & 3 & (c)  &  Thick fog (Visibility range $<$ 40 m)  & Slight curve (R=290~m)  & Issued  & Thick fog \\
    I & 4 & (b)  &  Thin fog (Visibility range $\geq$ 40 m)  & Slight curve (R=300~m)  & Not issued  & None \\
    I & 5 & (c)  &  Thick fog (Visibility range $<$ 40 m)  & Slight curve (R=310~m)  & Issued  & Thick fog \\
    I & 6 & (a)  &  Thin fog (Visibility range $\geq$ 40 m)  & Sharp curve (R=170~m)  & Issued  & Sharp curve \\
    I & 7 & (c)  &  Thick fog Visibility range $<$ 40 m)  & Slight curve (R=320~m)  & Issued  & Thick fog \\
    I & 8 & (b)  &  Thin fog (Visibility range $\geq$ 40 m)  & Slight curve (R=330~m)  & Not issued  & None \\
    I & 9 & (a)  &  Thin fog (Visibility range $\geq$ 40 m)  & Sharp curve (R=160~m)  & Issued  & Sharp curve \\
    II & 10 & (d)  &  Clear (Visibility range $\geq$ 40 m) & Sharp curve (R=150~m)  & Issued  & Sharp curve \\
    II & 11 & (e)  &  Thick fog Visibility range $<$ 40 m)  & Straight road  & Issued  & Thick fog \\
    III & 12 & (f)  &  Clear (Visibility range $\geq$ 40 m) & Slight curve (R=330~m)  & Not issued  & None \\ 
    III &  13 & (g)  &  Thin fog (Visibility range $\geq$ 40 m)  & Straight road  & Not issued  & None \\ \midrule
    IV & 14  & (h)  &  Clear (Visibility range $\geq$ 40 m) & Sharp curve (R=150~m)  & \begin{tabular}[r]{@{}r@{}}Not issued\\(RtI failure)\end{tabular}  & Sharp curve \\ 
    \bottomrule
  \end{tabular}
  \vspace{-2mm}
\end{table*}

\begin{figure*}
 \centering
  \begin{subfigure}[b]{0.245\linewidth}
  \centering
  \includegraphics[width=1\linewidth]{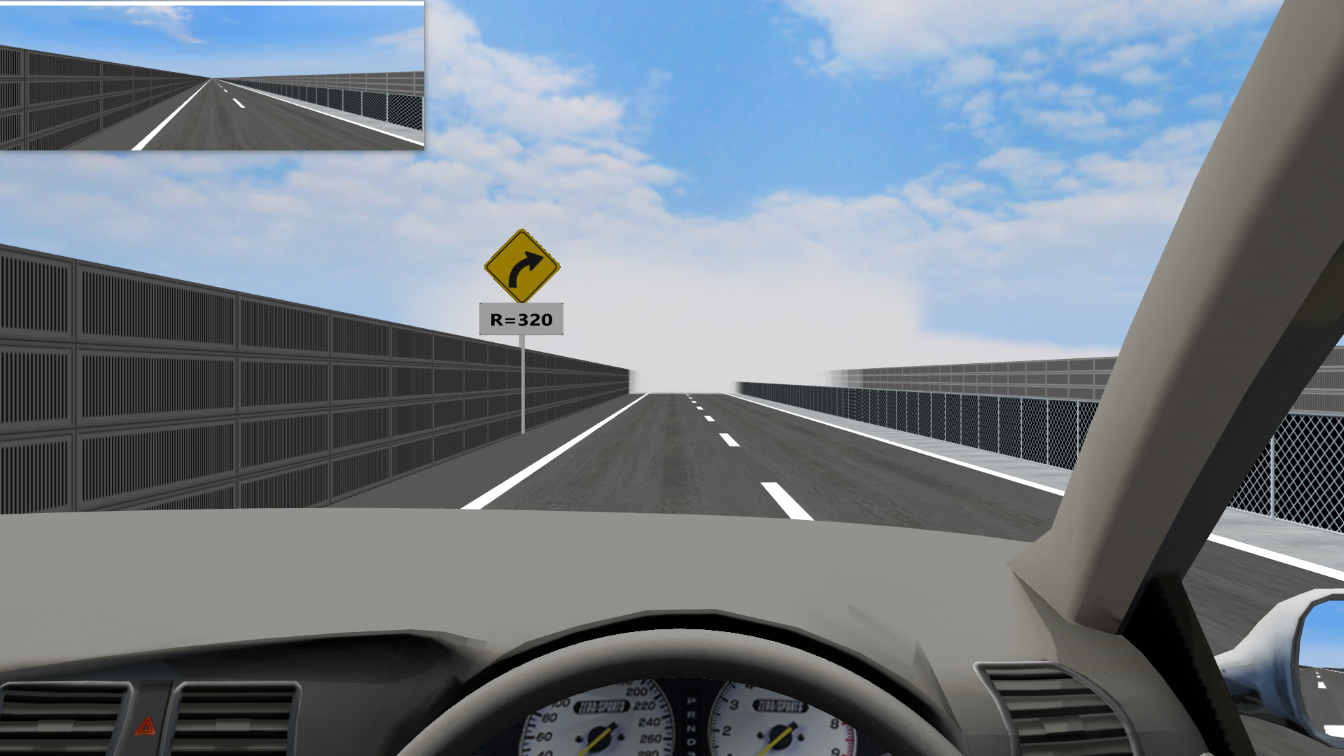}   
    \caption{Slight curve with thick fog.\\ \centering(Trial No.7)}
  \end{subfigure}
  \begin{subfigure}[b]{0.245\linewidth}
   \centering
  \includegraphics[width=1\linewidth]{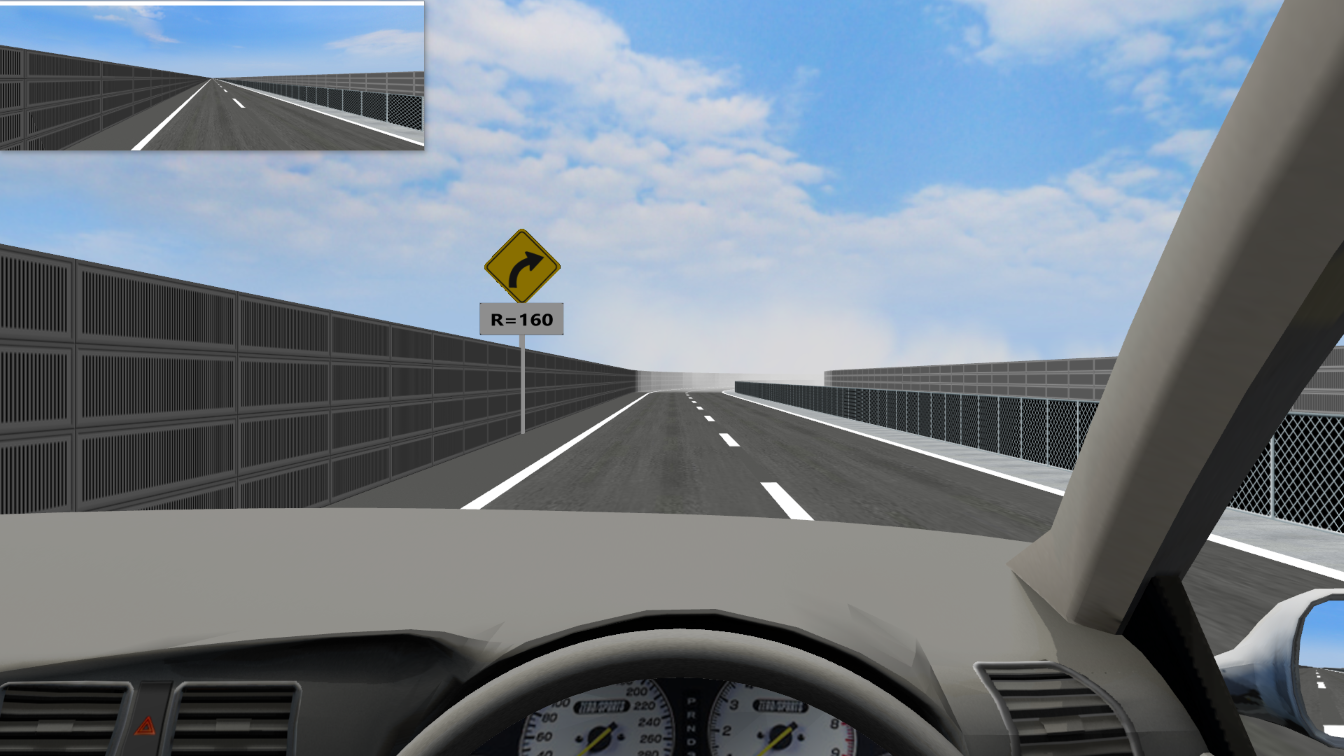}   
  \caption{Sharp curve with thin fog.\\ \centering(Trial No.9)}
  \end{subfigure}
  \begin{subfigure}[b]{0.245\linewidth}
 \centering
  \includegraphics[width=1\linewidth]{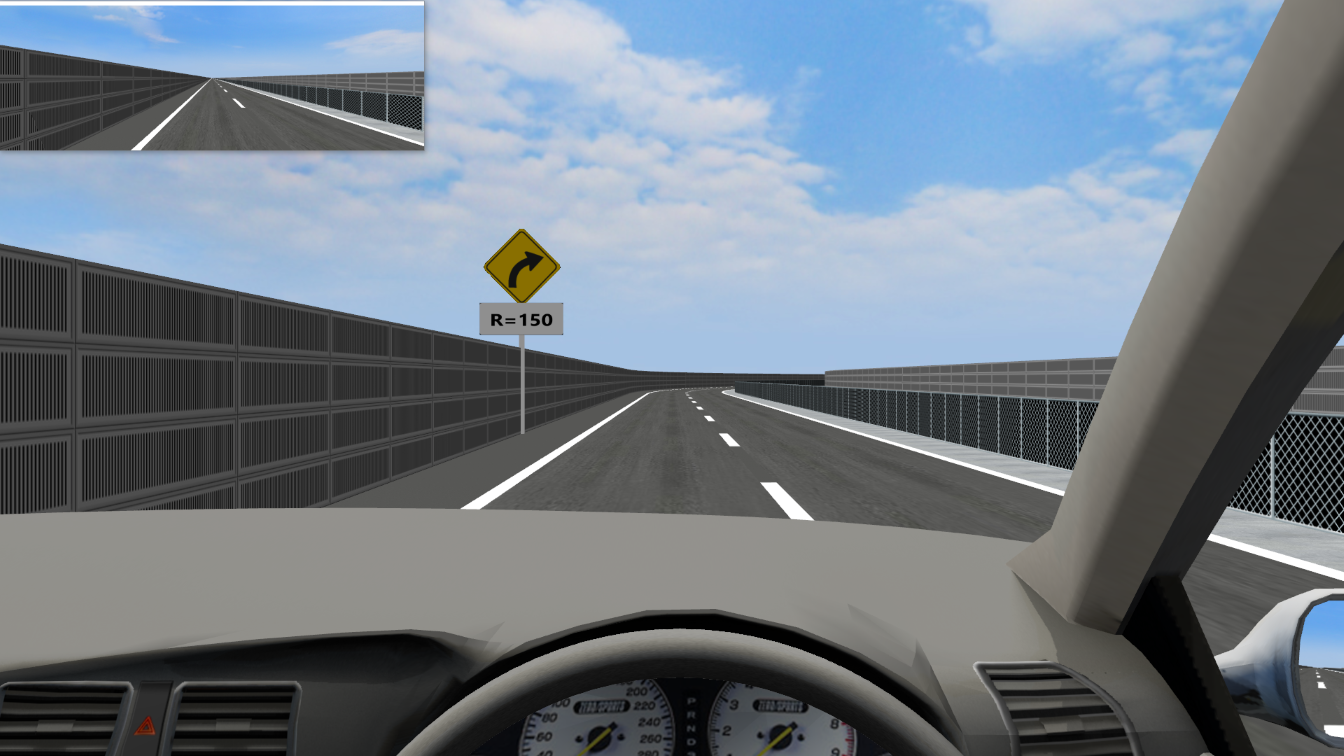}   
  \caption{Sharp curve without fog.\\  \centering(Trial No.10)}
  \end{subfigure}
   \begin{subfigure}[b]{0.245\linewidth}
  \centering
  \includegraphics[width=1\linewidth]{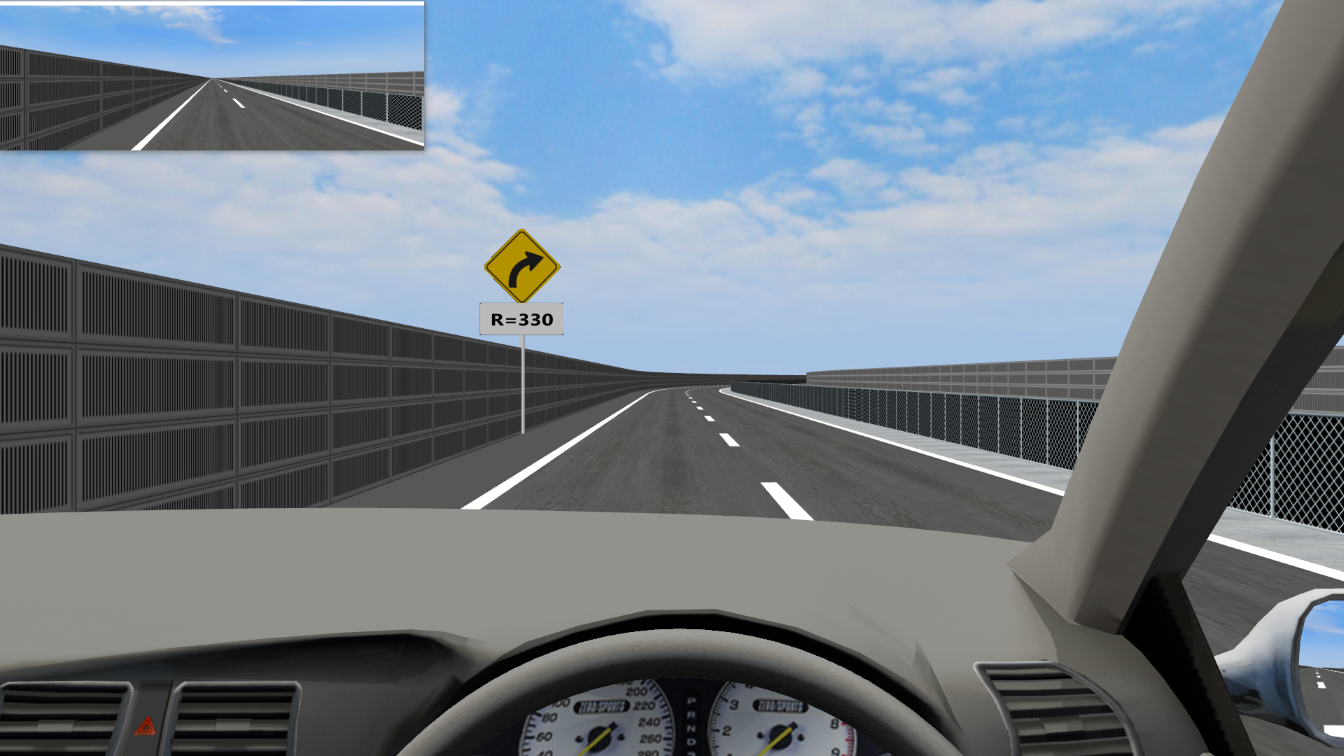}    
  \caption{Slight curve without fog.\\ \centering(Trial No.12)}
  \end{subfigure}
  
  \caption{Examples of the designed traffic scenarios as seen from the driver's view. (a) and (b) show thick and thin fog scenarios, respectively, while (c) and (d) show sharp and slight curves. Traffic signs show the curvature of curve roads.}
  \label{fig:scenario_from driver}
  \vspace{-3mm}
\end{figure*}

In this study, we focus on situations in which there appear to be multiple potential triggers contributing to RtI as perceived by drivers. 
Specifically, we are honing in on a specific scenario involving the simultaneous presence of fog and curves.
The reason for using these two factors is that in practical scenarios, the curvature information of the curve can be obtained in advance from the geographical information in the navigation system. 
Additionally, the density of fog can be recognized by camera~\cite{mori2007fog,cao2023fog} and millimeter wave radar~\cite{mori2007fog}.

As shown in Fig.~\ref{fig:scenario} and detailed in Table~\ref{table:Scenario}, we have meticulously crafted eight distinct driving scenarios tailored for employment on the two-lane highway. 
These particular scenarios have been meticulously selected to serve as the foundation for the four phases of the experiment.
To elaborate, the scenarios (a), (b) and (c) are used in the phase I, which appears to encompass multiple potential RtI triggers from the driver's perspective.
However, the ADS issues an RtI by only one correct RtI trigger in (a) and (c), respectively.
Moreover, the ADS did not issue RtI in scenario (b) because the visibility and road conditions did not exceed its system limitations.
In the experiment, we repeated these three scenarios three times to train the drivers to understand the system limitations of the ADS accurately.
In particular, when using the proposed RtI HMI to cue the correct RtI trigger to the driver, we assume that the driver can develop a better understanding of the system limitations by comparing multiple potential triggers with the correct trigger.
Table~\ref{table:Scenario} shows details of the visibility conditions and road conditions and the correct triggers for each of the nine scenarios (scenarios (a), (b), and (c) repeated three times).

Phase II comprises scenarios (d) and (e), each having only one potential RtI trigger, and both of them appropriately trigger the RtI.
In Phase III, scenarios (f) and (g) also have only one potential RtI trigger, separately, but those triggers can not activate the ADS RtI due to their visibility and road conditions (see Table~\ref{table:Scenario}).


Finally, in Phase IV, we designed an RtI failure scenario based on scenario (h). 
Unfortunately, as a result of a system failure, the driver is unable to receive pertinent information regarding the RtI before the ADS exceeds its system limitations and is forced to turn OFF.
Thus, the vehicle collides directly with the guardrail, if the driver can not take over the driving task actively.
In such a critical situation, only drivers who have a correct understanding of the limitations of the ADS will notice that something is wrong and take proactive action to take over vehicle control and avoid a collision. This highlights the importance of drivers being well-informed about the system's capabilities and limitations, allowing them to respond effectively to unforeseen circumstances.

Figure~\ref{fig:scenario_from driver} shows some examples of the designed traffic scenarios with different visibility and road conditions as seen from the driver's view. 
It is important to note that there is a traffic sign indicating the curvature of curve ahead based on the conditions of the Japan transportation system.
In addition, we placed the surrounding vehicles in front of the ego vehicle and the right lane to enhance the realism of the nine scenarios above. 
These surrounding vehicles maintain a safe following distance (more than 50~m) from the ego vehicle and travel at 80-100 km/h.
These vehicles do not suddenly decelerate or actively approach the experimental vehicle, posing a risk to the vehicle.
For the ego vehicle, we set the target speed of ACC to 80 km/h, and the maximum speed of the ego vehicle was limited to 110 km/h.

\subsection{Procedure}

First, the participants were introduced to the equipment and operation methods of the driving simulator, focusing on how to use the ADS to take control.
Subsequently, we conducted a pre-instruction regarding the system limitations of the ADS using an instruction manual.
This manual describes various conditions of system limitations, such as the set conditions of visibility and road curvature with their respective thresholds (see Section~\ref{sec:system_limitations}).
Additionally, other fake conditions of system limitation that did not occur in this experiment were also listed in the instruction manual to avoid making the set conditions conspicuous.
These fake conditions were took from the instruction manuals of vehicles with ACC and LKS currently available on the market, \eg road gradients, icy roads, tunnel entrances and exits.

After the pre-instruction, we asked the participants to practice using a driving simulator, such as manual driving, turning the ADS ON/OFF, and take-over.

The main experiment was conducted after the participants fully understood how to use the driving simulator.
Each participant experienced the 14 driving scenarios listed in Table~\ref{table:Scenario} and was requested to relax and look ahead during automated driving (subtask was not requested).
Furthermore, participants were required to respond to the RtI when it was issued.

After the driving experiment, a comprehension test regarding the ADS system limitations (see Section~\ref{sec:post-test}) was conducted.

\subsection{Measures}
We obtained the following three types of measurements during the experiment to answer the three research questions described in Section~\ref{sec:RQ},

\subsubsection{Post-experiment comprehension test}
\label{sec:post-test}

After completing all driving scenarios, we asked the participants to complete a comprehension test that presented a list of 13 scenarios (excluding the RtI failure scenario) from Table~\ref{table:Scenario}, along with their corresponding visibilities and road conditions specified by specific values (Table~\ref{table:Post-experiment comprehension test}).
We tasked the participants with determining whether the ADS could handle the automated driving tasks under the given conditions in these scenarios by three options: ``Yes,'' ``No'' or ``I don't know.''
One point was awarded for each correct answer, while no points were given for incorrect answers or selecting ``I don't know.''
Overall, we scored the comprehension test results from 0 to 13 points.

\begin{table}[t]
\footnotesize
\setlength\tabcolsep{5pt}
  \caption{Post-experiment comprehension test and its correct answers ($\bigcirc$).}  
  \label{table:Post-experiment comprehension test}
  \centering
  \scalebox{0.8}[0.8]{
  \begin{tabular}[H]{crcccc}
    \toprule
    No. & \begin{tabular}[c]{@{}c@{}}Visibility\\ conditions \end{tabular} & \begin{tabular}[c]{@{}c@{}}Road\\ conditions \end{tabular} & Yes & \begin{tabular}[c]{@{}c@{}}I don't\\know \end{tabular}& No \\
   \midrule
    1 & \begin{tabular}[c]{@{}c@{}}Thin fog (Visibility range $\geq$ 40 m) \end{tabular} & \begin{tabular}[c]{@{}c@{}}Sharp curve (R=180~m) \end{tabular} & & & $\bigcirc$ \\
    2 & \begin{tabular}[c]{@{}c@{}}Thin fog (Visibility range $\geq$ 40 m) \end{tabular} & \begin{tabular}[c]{@{}c@{}}Slight curve (R=280~m) \end{tabular} & $\bigcirc$ & & \\
    3 & \begin{tabular}[c]{@{}c@{}}Thick fog (Visibility range $<$ 40 m) \end{tabular} & \begin{tabular}[c]{@{}c@{}}Slight curve (R=290~m) \end{tabular} & & & $\bigcirc$ \\
    4 & \begin{tabular}[c]{@{}c@{}}Thin fog (Visibility range $\geq$ 40 m) \end{tabular}  & \begin{tabular}[c]{@{}c@{}}Slight curve (R=300~m) \end{tabular} & & & $\bigcirc$ \\
    5 & \begin{tabular}[c]{@{}c@{}}Thick fog (Visibility range $<$ 40 m) \end{tabular} & \begin{tabular}[c]{@{}c@{}}Slight curve (R=310~m) \end{tabular} & & & $\bigcirc$ \\
    6 & \begin{tabular}[c]{@{}c@{}}Thin fog (Visibility range $\geq$ 40 m) \end{tabular} & \begin{tabular}[c]{@{}c@{}}Sharp curve (R=170~m) \end{tabular} & & & $\bigcirc$ \\
    7 & \begin{tabular}[c]{@{}c@{}}Thick fog (Visibility range $<$ 40 m) \end{tabular}  & \begin{tabular}[c]{@{}c@{}}Slight curve (R=320~m) \end{tabular} & & & $\bigcirc$ \\
    8 & \begin{tabular}[c]{@{}c@{}}Thin fog (Visibility range $\geq$ 40 m) \end{tabular}  & \begin{tabular}[c]{@{}c@{}}Slight curve (R=330~m) \end{tabular} & $\bigcirc$ & & \\
    9 & \begin{tabular}[c]{@{}c@{}}Thin fog (Visibility range $\geq$ 40 m) \end{tabular}  & \begin{tabular}[c]{@{}c@{}}Sharp curve (R=160~m) \end{tabular} & & & $\bigcirc$ \\
    10 & \begin{tabular}[c]{@{}c@{}}Clear (Visibility range $\geq$ 40 m) \end{tabular} & \begin{tabular}[c]{@{}c@{}}Sharp curve (R=150~m) \end{tabular} & & & $\bigcirc$ \\
    11 & \begin{tabular}[c]{@{}c@{}}Thick fog (Visibility range $<$ 40 m) \end{tabular}  & Straight road  & & & $\bigcirc$ \\
    12 & \begin{tabular}[c]{@{}c@{}}Clear (Visibility range $\geq$ 40 m) \end{tabular} & \begin{tabular}[c]{@{}c@{}}Slight curve (R=330~m) \end{tabular} & $\bigcirc$ & & \\ 
    13 & \begin{tabular}[c]{@{}c@{}}Thin fog (Visibility range $\geq$ 40 m) \end{tabular}  & Straight road & $\bigcirc$ & & \\ 
    \bottomrule
  \end{tabular}
  }
 \vspace{-2mm}
\end{table}

\subsubsection{Take-over time to ADS OFF in the RtI failure scenario}

\begin{figure}[t]
  \centering
  \includegraphics[width=\linewidth]{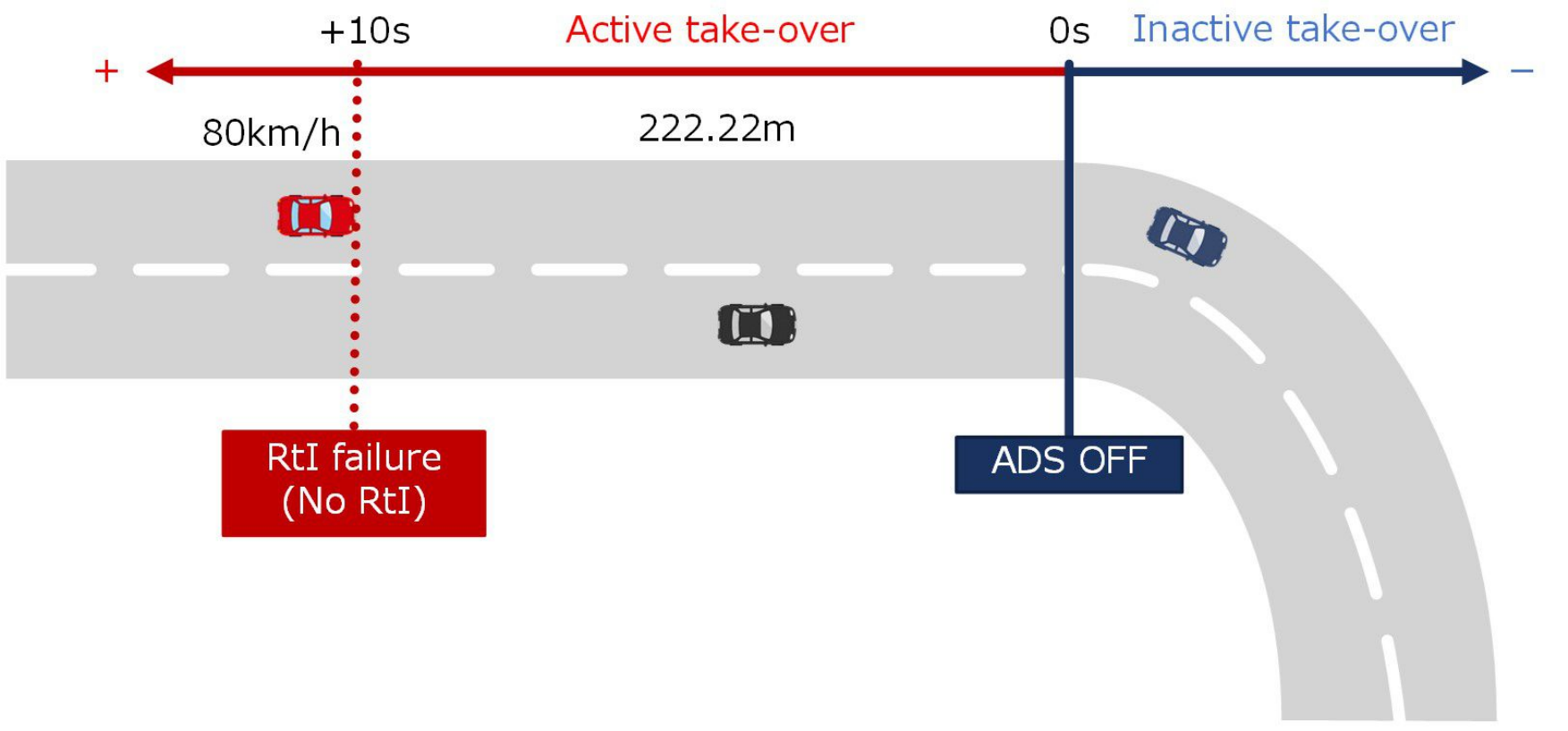}
  \caption{Illustration of an active take-over and an inactive take-over in the RtI failure scenario, \ie phase IV.}
  \label{fig:active_TO_image}
  \vspace{-6mm}
\end{figure}

Figure~\ref{fig:active_TO_image} shows an illustration of an active take-over and an inactive take-over in the RtI failure scenario, \ie phase IV.
The time points at which the takeover occurred by two groups of participants in the RtI failure scenario were measured.
We set the time point at which the timing of ADS disengagement (ADS OFF) as the 0-point (0~[s]).
The time before the 0-point are defined as positive values, while times after the 0-point are defined as negative values.
Furthermore, if the driver takes over before the 0-point, it is classified as an active take-over.
On the other hand, if the driver takes over after the 0-point, it is considered an inactive take-over, which significantly increases the risk of collision with the guardrail.

\subsubsection{Number of collisions in the RtI failure scenario}

We counted the number of participants who collided in the RtI failure scenario for two groups, \ie the trigger cue \& reason group and \textit{w/o trigger cue} group (see Section~\ref{sec:participants}), separately.

\section{RESULTS}

\subsection{Post-experiment comprehension test}

The results of the post-experiment comprehension is shown in Fig.~\ref{fig:Test}.
The results indicate that the median test score for the group \textit{w/ trigger cue \& reason} and \textit{w/o trigger cue} groups was 13 and 8 points, respectively. 
The Mann–Whitney U test ($U=6.000$, $p=0.000$, CLES $=0.060$, two-sided) suggests that the test scores of the \textit{w/ trigger cue \& reason} group were significantly higher than those of the \textit{w/o trigger cue} group.

\subsection{Take-over time to ADS OFF in the RtI failure scenario}

Figure~\ref{fig:Take-over time to ADS OFF} presents the results of the take-over time to ADS OFF in RtI failure scenario.
T-test ($T=-2.948$, $p=0.009$, cohen-d $=1.318$, two-sided) suggests that the take-over time of the \textit{w/ trigger cue \& reason} group were significantly higher than those of the \textit{w/o trigger cue} group.

Figure~\ref{fig:Number of participants whose take-over time} presents the results of the number of participants whose take-over time to ADS OFF in the RtI failure scenario.
These results suggest that the number of participants of the \textit{w/ trigger cue \& reason} group whose take-over time were higher than those of the \textit{w/o trigger cue} group.

\subsection{Number of collisions in RtI failure scenario}

\begin{figure}[htp]
  \centering
  \includegraphics[width=0.7\linewidth]{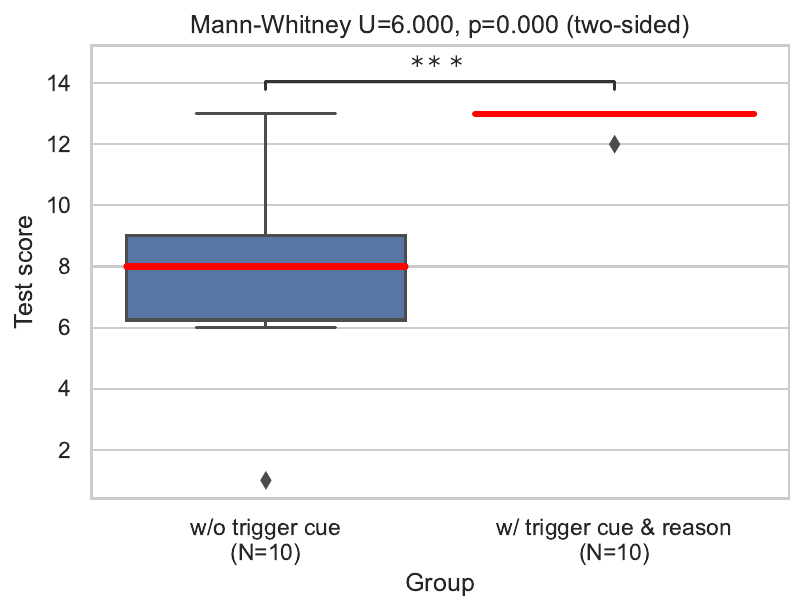}
  \caption{Post-experiment comprehension test results of two participant groups.}
  \label{fig:Test}
    \vspace{3mm}
 \centering
  \includegraphics[width=0.7\linewidth]{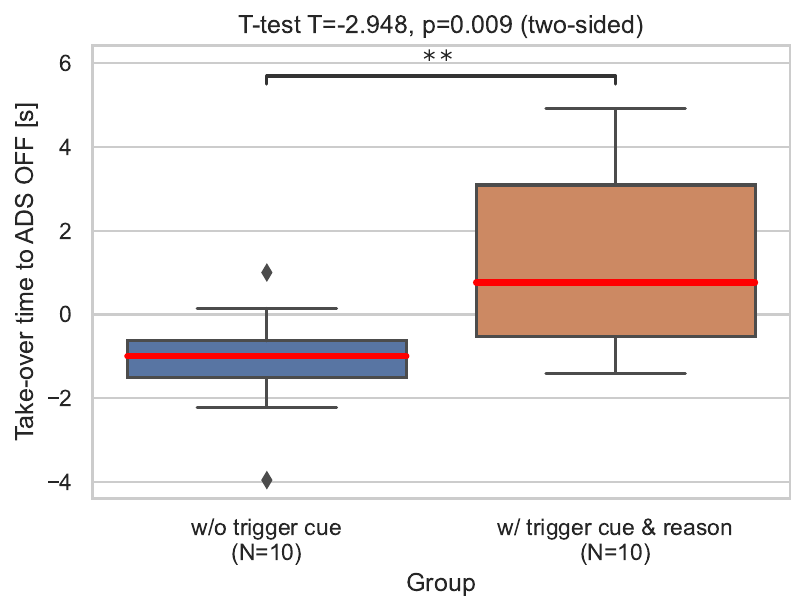}
  \caption{Take-over time to ADS OFF in the RtI failure scenario, \ie phase IV.}
  \label{fig:Take-over time to ADS OFF}
  \vspace{3mm}
 \centering
  \includegraphics[width=0.7\linewidth]{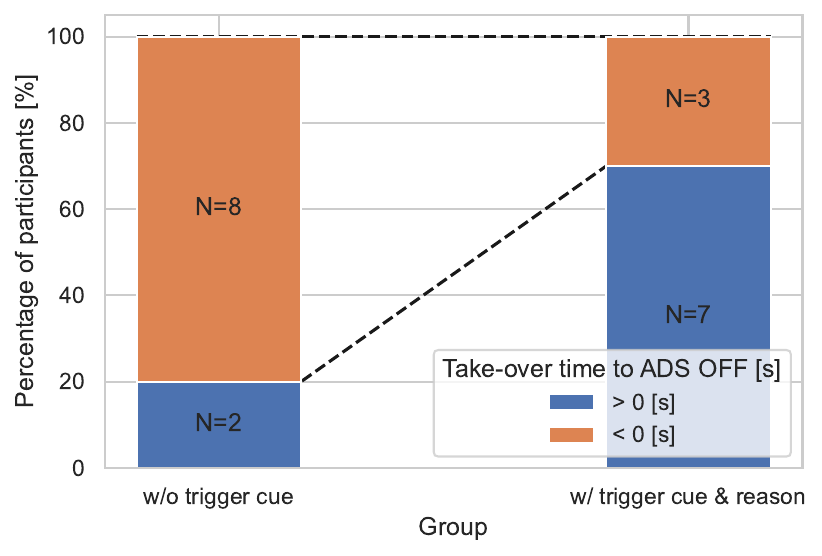}
  \caption{Number of participants who actively takeover in the RtI failure scenario, \ie phase IV.}
  \label{fig:Number of participants whose take-over time}
  \vspace{3mm}
 \centering
  \includegraphics[width=0.7\linewidth]{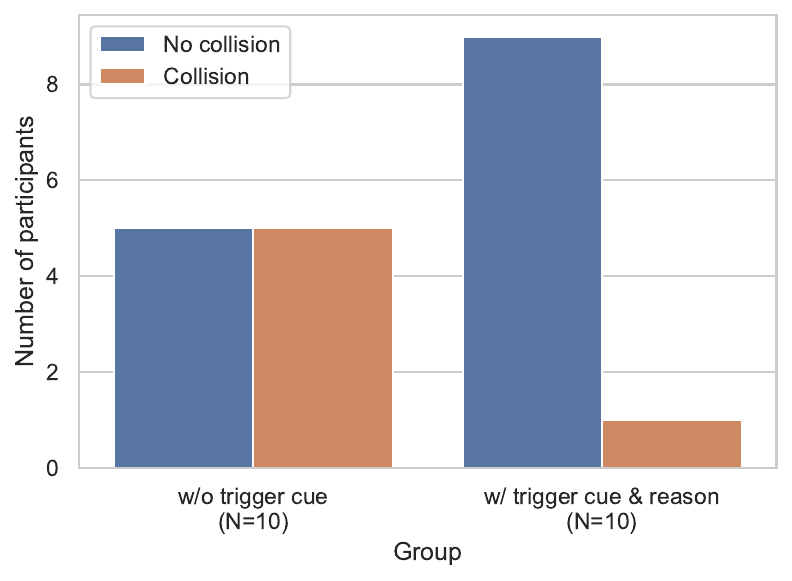}
  \caption{Number of participants in two groups who had a collision in the RtI failure scenario, \ie phase IV.}
  \label{fig:collision}
  \end{figure}

Figure~\ref{fig:collision} presents the number of collisions in the RtI failure scenario.
Specifically, in the \textit{w/ trigger cue \& reason} group, nine participants did not collide, but one had.
In the \textit{w/o trigger cue} group, five had no collisions, while five had.
Fisher's exact test suggested that there was no significant difference between the number of collisions between two groups ($p=0.141$).

\section{DISCUSSION}

\subsection{Post-experiment comprehension test}

In response to RQ~1, the comprehension test presented in Fig.~\ref{fig:Test} indicates that the test scores of the group that had the trigger cue \& reason on RtI were significantly higher than those of the group that did not get the trigger cues.
The results suggest that drivers will correctly understand the system limitations of ADS through repeated use of the proposed HMI.
In addition, the high scores on the post-experiment comprehension test for the \textit{w/ trigger cue \& reason} group also suggest that the proposed RtI HMI had a sustained educational effect compared to the pre-educational method (i.e., the pre-instructions received by the \textit{w/o trigger cue} group).
This finding aligns with the conclusion in~\cite{merriman2023does} which showed that the driver's training affects their mental models, \ie their comprehension of system limitations.

\subsection{Take-over time to ADS OFF in RtI failure scenario}

In response to RQ~2, the take-over times to ADS OFF in the RtI failure scenario in Fig.~\ref{fig:Take-over time to ADS OFF} indicate that the take-over times of the \textit{w/ trigger cue \& reason} group were significantly higher than those of the \textit{w/o trigger cue} group.
In addition, the number of participants whose take-over time in the RtI failure scenario in Fig.~\ref{fig:Number of participants whose take-over time} indicates that the take-over time of the group that had the trigger cue \& reason on RtI were higher than those of the group that did not get the trigger cues.
These results suggest that drivers using the proposed HMI actively take over driving in hazardous situations.
These findings align with the conclusion in~\cite{zhou2020effect} which showed that the driver's comprehension of system limitation affects the take-over time in the unexpected hazards due to system failures.

\subsection{Number of collisions in RtI failure scenario}

Figure~\ref{fig:collision} shows that the number of collisions in the RtI failure scenario was lower in the \textit{w/ trigger cue \& reason} group than in the \textit{w/o trigger cue} group.
These results suggest that the collision risk was reduced when the driver was provided with the trigger cue \& reason in the RtI. 

Combining the comprehension test results presented in Fig.~\ref{fig:Test}, we can conclude RQ~3 that the results imply that the presentation of trigger cues \& reason might help drivers understand the system limitations and reduce the risk of collisions.
This observation is in harmony with the findings presented in previous studies such as~\cite{zhou2021effects, zhou2021influence}, wherein it was demonstrated that a driver's awareness of system limitations significantly influences takeover performance.
Moreover, using trigger cues associated with weather and road conditions aligns with~\cite{wright2018effective} which showed that environmental cues increase the likelihood of a driver mitigating a collision.

\subsection{Limitations}

The number of participants, which was limited to 10 in each group, may have constrained the statistical significance of the results.
We designed a driving simulator experiment with more participants to address this limitation and obtain robust and reliable results.


The order effect of the scenarios is another limitation. 
As Table.~\ref{table:Scenario} shows, we intentionally fixed the order of the scenes to focus on changes caused by the effect of the trigger cue. 
In particular, the learning effects of RtI triggers may vary depending on the order of the scenarios.

\section{CONCLUSION}

We proposed a voice-based HMI, \ie RtI trigger cues \& reason, to help drivers correctly understand the system limitation of an ADS in scenarios with multiple potential RtI triggers.
The results of a between-group experiment using a driving simulator showed that incorporating effective trigger cues \& reason into the RtI enabled drivers to comprehend the ADS's system limitations better and reduce collisions.
Furthermore, the use of the proposed RtI HMI could also encourage drivers to actively stay informed about the situation and takeover the control in hazardous conditions.
Therefore, the proposed method is important to promote the safe use of ADS during the takeover process.

In future work, we will increase the number of participants and further analyze the impact of RtI cues and reason on their behavior during take-overs, such as reaction time and operational behavior.
We will also observe whether RtI leads to drivers developing over-trust in ADS and explore methods to mitigate this issue.

\section*{ACKNOWLEDGMENTS}
This work was supported by JSPS KAKENHI Grant Numbers 20K19846 and 22H00246, Japan.

\bibliographystyle{IEEEtran} 

\bibliography{sample.bib}

\end{document}